\documentstyle[aps,psfig,12pt]{revtex}
\def\beq{\begin{equation}}
\def\eeq{\end{equation}}

\def\bea{\arraycolsep .1em \begin{eqnarray}}
\def\eea{\end{eqnarray}}
\def\Tr{{\rm Tr}}

\let\Ga=\Gamma

\def\eq#1{(\ref{#1})}

\def\s0#1#2{\mbox{\small{$ \frac{#1}{#2} $}}}
\def\0#1#2{\frac{#1}{#2}}

\begin{document}
\begin{center}

\thispagestyle{empty}

{\normalsize\begin{flushright}HD-THEP-00-28\\[12ex] 
\end{flushright}}

\mbox{\large \bf Optimisation of the exact renormalisation group} \\[6ex]

{Daniel F. Litim
\footnote{D.Litim@thphys.uni-heidelberg.de}
}
\\[4ex]
{\it Institut f\"ur Theoretische Physik\\ Philosophenweg 16, 
D-69120 Heidelberg, Germany.}
\\[10ex]
 
{\small \bf Abstract}\\[2ex]
\begin{minipage}{14cm}{\small
A simple criterion to optimise coarse-grainings for exact renormalisation group equations is given. It is aimed at improving the convergence of approximate solutions of flow equations. The optimisation criterion is generic, as it refers only to the coarse-grained propagator at vanishing field. In physical terms, it is understood as an optimisation condition for amplitude expansions. Alternatively, it can be interpreted as the requirement to move poles of threshold functions away from the physical region. The link to expansions in field amplitudes is discussed as well. Optimal parameters are given explicitly for a variety of different coarse-grainings. As a by-product it is found that the sharp cut-off regulator does not belong to the class of such optimal coarse-grainings, which explains the poor convergence of amplitude expansions based on it.}
\end{minipage}
\end{center}

\newpage
\pagestyle{plain}
\setcounter{page}{1}
\noindent
{\bf Introduction}\\[-1ex]

Flow equations \cite{Wilson} or exact renormalisation group equations are a promising tool for non-perturbative problems within quantum field theories. Based on the idea of integrating-out large momentum modes within a path integral formulation of quantum field theory, they permit to obtain an effective (or `coarse-grained') theory for the remaining light degrees of freedom. Flows interpolate between some initial (classical) action, and the full (quantum) effective action. The particular strength of the formalism is its flexibility, allowing for systematic approximations without being tied to the small coupling region \cite{Wilson,CW}.
 
On the conceptual side, important progress has been made within the recent past, in particular through the consistent extension to (non-Abelian) gauge theories \cite{gaugefields}, which may play an important role in future applications to QCD. In turn, comparatively little attention has been paid to the interplay between {\it approximate} solutions to flow equations on the one side and the coarse-graining procedure on the other. Any explicit application of the formalism requires some approximation, simply because it seems very difficult to simultaneously solve infinitely many coupled partial differential equations. {Approximate} solutions have a finite domain of validity \cite{LitimTetradis}, they may show a spurious dependence on the coarse-graining parameters at second order phase transitions \cite{Ball:1995ji} or first order ones \cite{Litim96,Freire:2000sx}, and the corresponding expansions have a finite (often unknown) radius of convergence \cite{Morris:1999ba}. It has also been argued that the spurious scheme dependence can be used to assess the reliability of a given approximation \cite{Litim96}, resulting in error bars describing the scheme and truncation dependence \cite{Freire:2000sx}, or to obtain a better convergence employing a minimum sensitivity condition \cite{LPS}. 
 
The cross-dependence between the coarse-graining and a truncation of the flow has a very simple origin. Any coarse-graining couples to all operators in the effective action, with a strength depending on the details of the specific implementation. The sharp cutoff, for example, eliminates all momentum modes above some coarse-graining scale $k$, while smooth cutoff functions, like exponential, algebraic or mass-like ones, smoothly cut off the higher modes within some finite momentum interval about $k$. Hence, a change of the coarse-graining scheme alters the effective coupling to the operators in the effective action. A strong scheme dependence indicates whether some relevant operators have been neglected within a given approximation \cite{Freire:2000sx}. Along similar lines it has also been argued that the smoothness of the coarse-graining can influence the contributions of higher order derivative operators \cite{LPS}. 

More generally, this interplay suggests that a suitably chosen coarse-graining can significantly improve the convergence properties or the domain of validity of approximate solutions, which  would be of great value for both analytical and numerical applications. In the present note we take profit of this freedom to propose a simple yet efficient optimisation criterion for Wilsonian flows, with the following characteristics: it optimises approximate solutions to flow equations; it is compatible with systematic approximation schemes like the derivative expansion; it only relies on a basic ingredient to the flow equation (the effective coarse-grained propagator at vanishing field); it is generically applicable, {\it i.e.}~not based on a specific theory. 
\\[3ex]
\noindent
{\bf Flows, coarse-grainings and propagators}\\[-1ex]

To begin with, we briefly review the basic ingredients needed for a Wilsonian flow equation. The modern way to implement an exact renormalisation group procedure amounts to add a regulator term $\sim \int d^dq \phi(-q) R_k(q^2)\phi(q)$ (for bosonic fields) to the action \cite{CW}. The operator $R_k(q^2)$, sometimes referred to as the regulator scheme (RS), desribes the coarse-graining and introduces a fiducial scale parameter $k$, which is ultimately interpreted as a coarse-graining scale. This scale parameter induces a scale dependence, which, when written for the scale-dependent effective action $\Ga_k$, results in the flow equation
\beq \label{general}
\0{\partial}{{\partial t}}\Ga_k[\phi]
=\012\Tr\left\{\left(\Ga^{(2)}_k[\phi]+
 R_k\right)^{-1}\0{\partial R_k}{{\partial t}}\right\}. 
\eeq
Here,  $\phi$  denotes bosonic fields and  $t = \ln k$ the logarithmic scale parameter.  The right hand side of \eq{general}  contains the regulator function $R_k$ and the second functional derivative of the effective action with respect to the fields. The trace denotes a summation over all indices and integration over all momenta. 

All specifications regarding the coarse-graining are given through the operator $R_k(q^2)$. This operator can be chosen at will though within some basic restrictions which are briefly reviewed. First of all, it is required that $R_k(q^2\to 0)>0$. This ensures that the effective propagator at vanishing field $\Delta_k(q^2)=1/[{q^2+R_k(q^2)}]$ remains finite in the infrared limit $q^2\to 0$, and no infrared divergences are encountered in the presence of massless modes. The second requirement is the vanishing of $R_k$ in the infrared, $R_k(q^2)\to 0$ for $k\to 0$. This guarantees that the coarse-grained generating functional of 1PI Green functions $\Gamma_k$ reduces to the usual generating functional of 1PI Green functions $\Ga=\lim_{k\to 0}\Ga_k$. The third condition to be met is that $R_k(q^2)$ diverges in the UV limit $k\to\Lambda$. This way it is ensured that the microscopic action $S=\lim_{k\to \Lambda}\Ga_k$ is approached in the ultraviolet limit $k\to \Lambda$. These conditions ensure that the flow \eq{general} interpolates between the classical and the quantum effective action. For later convenience, we shall impose a further constraint on $R_k$, which, however, is nothing but a normalisation condition for the coarse-graining scale $k$, namely $R_k(q^2=k^2)=k^2$. If one would have chosen a regulator with  $R_k(q^2=k^2)=c\,k^2$, then it is always possible to find a corresponding scale $k_{\rm eff}=f(c)k$ such that $R(q^2=k^2_{\rm eff})=k^2_{\rm eff}$. Hence, the last condition guarantees that (trivial) rescalings of $k$ have been factored-out, and comparing different effective propagators is now sensible. 
 
It is important to realise that the integrand of the flow equation \eq{general}, as a function of momenta $q$, is peaked about $q^2\approx k^2$, and suppressed for large momenta. Consequently, at each infinitesimal integration step $k\to k-\Delta k$ only a narrow window of momentum modes contribute to the change of $\Ga_k\to \Ga_{k-\Delta k}$. Most importantly, modes with momenta $q\gg k$ no longer contribute to the running at the scale $k$. It is this property which justifies the interpretation of $\Gamma_k$ as a coarse-grained effective action with modes $q\gg k$ already integrated out. 
 
Let us now turn to the effective coarse-grained propagator $\Delta_k(q^2)$, or, for convenience, to its dimensionless inverse given by
\beq\label{P2}
P^2\equiv 1/k^2\Delta_k(q^2)=y[1+r(y)]\ ,
\eeq
and $y=q^2/k^2$. We find it convenient to write the regulator function $R_k(q^2)$ in terms of a dimensionless function $r(q^2/k^2)$ as
\beq
R_k(q^2)=q^2\, r(q^2/k^2)\ ,
\eeq
normalised as   $r(1)=1$. As a consequence of the above conditions, we can establish the following properties of $P^2$ (see Fig.~1). In contrast to the `free' inverse propagator $q^2$, the coarse-grained one is strictly positive for all momenta , $P^2>0$. This is due to the regulator term. For large momenta , $P^2$ grows linearly with $y=q^2/k^2$. 

\begin{center}
\begin{tabular}{ccc}
\hline\hline\\[-1ex]
$\quad{}$Class$\quad{}$ 
& $\quad{}P^2(q^2=0)\quad{}$ 
& $\min_{q^2} P^2$ at \\[.5ex] \hline\\[-1.5ex] 
Ia:    & finite     & $q^2 =    0$ \\
Ib:    & finite     & $q^2 \neq 0$ \\
II:    & infinite   & $q^2 \neq 0$ \\[1ex]
\hline\hline
\end{tabular}
\end{center}
\begin{center}
\begin{minipage}{.55\hsize}
\vskip.3cm
{\label{Tab1}\small Tab.~1: Classification of different regulator schemes.\\ }
\end{minipage}
\end{center}
For small momenta, $r(y)$ diverges at least as $1/y$, hence  $P^2$ either approaches a constant $>0$, or it diverges.  We refer to the first class of regulators as mass-like, or Class I regulators, and to the second one as Class II regulators (see Tab.~1). Notice that their proper normalisation implies $P^2(q^2=k^2)=2$ for any regulator.
\begin{figure}[t]
\begin{center}
\unitlength0.001\hsize
\begin{picture}(500,500)
\put(180,350){\framebox{\large $P^2\left({q^2}/{k^2}\right)$}}
\put(92,350){\Large II}
\put(65,268){\Large Ib}
\put(55,152){\Large Ia}
\put(200,-20){\large ${q^2}/{k^2}$}
\psfig{file=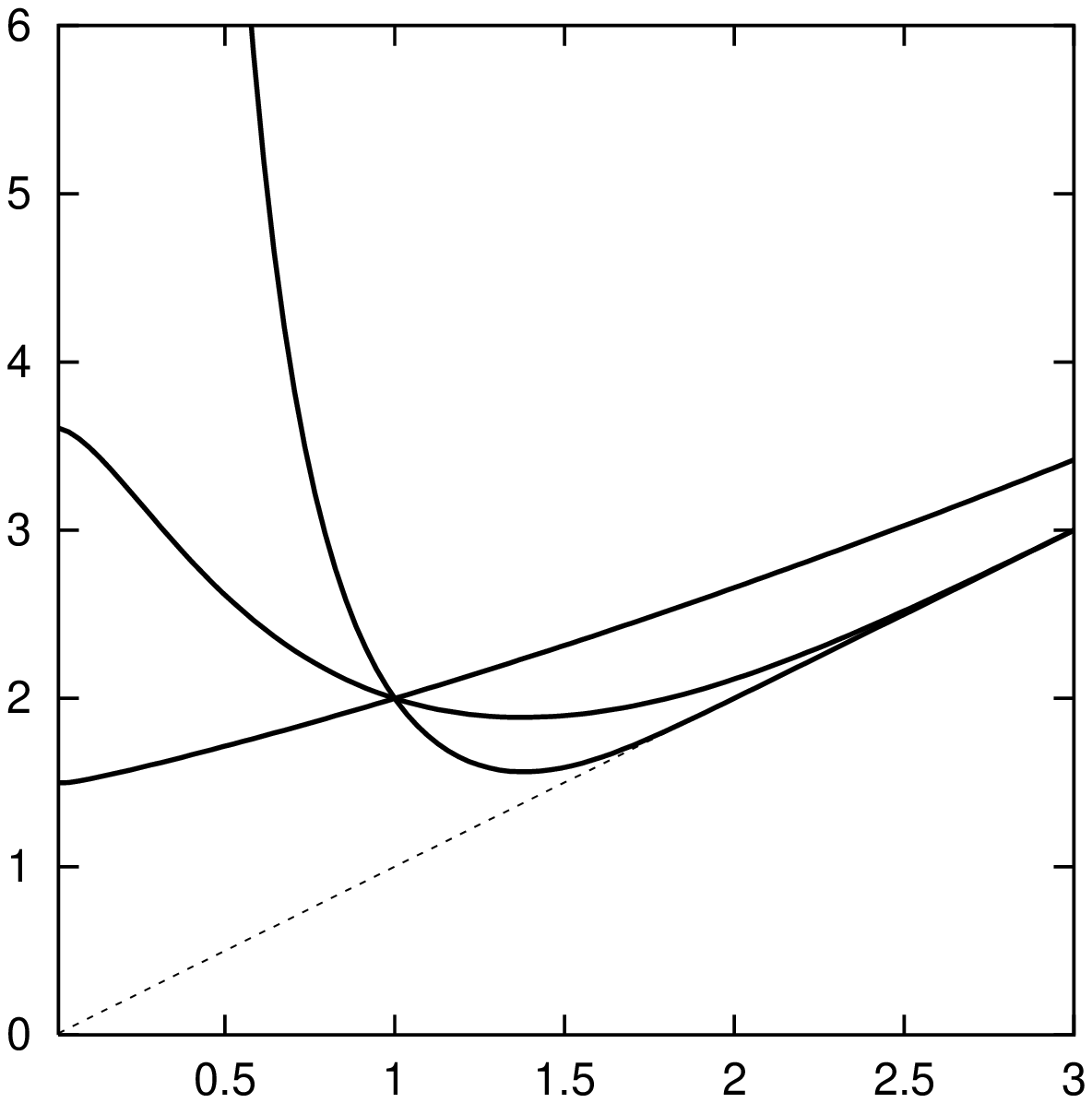,width=.45\hsize}
\end{picture}
\vskip.8cm
\begin{minipage}{.92\hsize}
{\small Fig.~1: The effective inverse propagator for the different regulator classes Ia, Ib and II (thick full lines). The physical normalisation implies that $P^2(y=1)=2$ for all regulators. The dashed line corresponds to $r=0$.}
\end{minipage} 
\end{center}
\end{figure}
\noindent
{\bf Optimisation}\\[-1ex]

The effective coarse-grained inverse propagator $P^2$ depends on the particular regularisation scheme chosen. We can formulate a simple optimisation criterion, which is the requirement that the minimum of $P^2$ with respect to momenta be maximal with respect to the coarse-graining,
\beq\label{opt}
C_{\rm opt}=\max_{\rm RS}\left(\min_{y\ge 0} P^2(y)\right)\equiv
\max_{\rm RS}\left(\min_{q^2\ge 0}\left[k^2\Delta_k(q^2)\right]^{-1}\right)\ .
\eeq
Here, we denote those coarse-grainings as `optimal' for which the maximum is attained. It is interesting to note that this criterion is based only on the effective propagator at vanishing field, that is, on the basic ingredient to the flow equation \eq{general}. Stated differently, no reference is made to a specific model or theory considered. This is desirable insofar as it renders the optimisation condition universally applicable. 
 
Some few observations concerning $C_{\rm opt}$ can be made at this stage. For an optimal scheme, we have $C_{\rm opt}=2$. This follows from combining $r(1)=1$ with \eq{opt}, which implies $\min_y P^2(y)\le P^2(y=1)=2$ for any scheme. This also implies that the optimal $P^2$ reaches its (local) minimum at $y=1$ and $r=1$ for any optimised scheme. Such regulators are either of Class Ib or Class II. A Class Ia regulator can only be optimal in the event where $P^2(y)$ has two degenerate minima, one at $y=0$ (by definition), and the other one at $y=1$. Hence, an optimal Class Ia regulator is also of Class Ib. The optimal value $C_{\rm opt}=2$ cannot be reached for regulators, which are strictly Class Ia ($P^2(y=0)$ is the global minimum).
\\[3ex]
\noindent
{\bf Threshold functions} \\[-1ex]

The optimisation criterion can be interpreted in more physical terms when generic threshold functions are considered. To give a particular example, consider a $N$-component real scalar field theory in $d$ dimensions. To leading order in the derivative expansion, we approximate the effective action as $\Gamma_k=\int d^dx[\s012\partial_\mu\phi^a\partial^\mu\phi_a+U_k(\phi)+O(\partial^4)]$ and $a=1,\ldots,N$. It is useful to introduce rescaled (dimensionless) variables as  $u(\rho)=U_k/k^d$ and $\rho=\s012\phi^a\phi_a k^{2-d}$. The flow equation for the first derivative of the effective potential $u'$ is a second-order partial differential equation, to wit
\beq\label{example}
\partial_t u'=-2u'+(d-2)\rho u''
-2v_d(N-1)u''\,\ell^d_1(\omega_1)-2v_d(3u''+2\rho u''')\ell^d_1(\omega_2)
\eeq
with $v^{-1}_d=2^{d+1}\pi^{d/2}\Gamma[\0d2]$. Here, the relevant dimensionless amplitudes are $\omega_1\equiv u'$ and $\omega_2\equiv u'+2\rho u''$, and the threshold function $\ell^d_1(\omega)$ reads
\beq\label{thresholdexample}
-\int_0^\infty dy\,
r'(y)\, y^{1+d/2}\01{(P^2+\omega)^2}\ .
\eeq
Flow equations of the form \eq{example} have been intensively studied in the literature \cite{Tetradis:1994ts,MorrisSharp,AbelHiggs3d,LitimTetradis,Freire:2000sx}.
 
In the general case, an Ansatz for $\Gamma_k$ need not be restricted to the leading order in the derivative expansion. At higher order, one finds coupled sets of partial differential equations for the coefficient functions of the higher derivative operators, involving more complicated threshold functions. They are obtained from the flow equation \eq{general}, or functional derivatives thereof, within specific models and approximations. A generic threshold function has the following structure,
\beq\label{threshold}
\int_0^\infty dy \,K_{n_1,\ldots,n_i}(y) \,[P^2(y)+\omega_1]^{-n_1}\ldots [P^2(y)+\omega_i]^{-n_i}\ 
\eeq 
and all $n_i>0$. Typically, the amplitudes $w_i$ describe (field-dependent) mass terms $\omega\sim m_k^2(\phi)/k^2$, which grow large once the coarse-graining scale falls below the mass threshold, $k\ll m_k(\phi)$. This leads to the decoupling of the corresponding modes (hence the name `threshold' functions). In dimensionful units the physical amplitudes are $\Omega\equiv k^2\omega\sim m_k^2(\phi)$. The explicit form of the kernel $K_{n_1,\ldots,n_i}$ depends on the specific quantity under investigation. In the example above, the threshold function \eq{thresholdexample} is of the form \eq{threshold} for the specific kernel $K_2(y)=-r'(y)\, y^{1+d/2}$.
 
As an immediate consequence of the properties of the basic flow equation \eq{general} we can conclude that the kernel of \eq{threshold} is well-defined, finite and peaked as a function of momenta. Given the properties of $P^2$, we deduce that such threshold functions always have a pole on the negative $\omega$-axis, located at $\omega_{\rm pole}= -\min_{y\ge 0} P^2(y)<0$. The location of the pole depends on the particular coarse-graining chosen. Notice that the pole cannot be shifted arbitrarily far away towards the negative axis. Requiring that the pole of any threshold function be as negative as possible corresponds precisely to the condition \eq{opt}. Hence, the maximally attainable point is given by $C_{\rm opt}\equiv -\omega_{\rm pole}=2$. The condition \eq{opt} applies to any threshold function of the form \eq{threshold}, and therefore holds for any amplitude $\omega>\omega_{\rm pole}$. In dimensionful units, the pole is located at $\Omega_{\rm pole}=k^2\omega_{\rm pole}<0$. For $k$ close to some ultraviolet scale $\Lambda$ the pole is far away on the negative axis and outside the physical domain. It is moving inward for decreasing $k$, ultimately reaching $\Omega_{\rm pole}=0$ in the infrared limit, where the potential becomes convex.
 
The pole structure of threshold functions is closely linked to the flattening of the non-convex parts of effective potentials in the limit $k\to 0$, like $U_k$ in the example above \cite{Tetradis:1992qt}. The convexity is a consequence of the 1PI generating functional being obtained by a Legendre transformation. For finite $k>0$, though, convexity is not required. In the non-convex region, the amplitudes $\omega_1\sim k^{-2}\phi^{-1}\partial U_k/\partial \phi$ or $\omega_2\sim k^{-2}\partial^2 U_k/\partial \phi^2$ turn negative. In particular, for $k\to 0$, they approach the poles of threshold functions. Hence, choosing a coarse-graining for which this pole is the most negative corresponds to choosing a coarse-graining for which the transition towards a convex effective action is the smoothest. Such a choice is particularly useful for the stability and an improved convergence of numerical implementations of flow equations.
\\[3ex]
\noindent
{\bf Amplitude expansions}\\[-1ex]

An alternative physical explanation of the optimisation condition is given in terms of an expansion in (small) amplitudes. With amplitude expansion we mean any expansion in the amplitudes $\omega$ in units of $P^2$. Such expansions {\it always} exist, simply because $P^2$ is strictly non-zero. Then, all information about the regulator scheme is encoded in the related expansion coefficients, introduced in \cite{Litim96}. Here, they can be written as
\beq\label{ak}
a_n=\int_0^\infty dy\, K[r]\, P^{-n}(y)\ .
\eeq 
The explicit form of the kernel $K$ depends on the specific quantity studied. For the example given above, the threshold function is Taylor-expanded in powers of the amplitudes $\ell^d_1(\omega)=\sum_{m=1}^{\infty}\0{2m}{d}\,a_{2m-d}\,(-\omega)^{m-1}$ where the coefficients $a_n$ are those defined in \eq{ak} with the particular kernel
\beq\label{Kpot}
K[r]=-\0d2 \0{r'(y)}{[1+r(y)]^{1+d/2}}\ ,
\eeq
normalised such that $a_0=1$ \cite{Litim96}. 
 
The main characteristics of amplitude expansions are deduced from the knowledge that the kernel $K$ for general expansion coefficients \eq{ak} is finite, peaked, and suppressed for sufficiently large momenta $q^2/k^2$. Conceptually speaking, it would be most desirable if only a few terms in such a series have to be taken into account, and higher order corrections be small. Hence, we look for coarse-grainings for which the main physical information is already encoded within a few leading order terms. For fixed amplitudes, this amounts to the requirement that the higher order coefficients $a_n$ be as small as possible, or, that the radius of convergence for amplitude expansions be maximal. The (dimensionless) radius of convergence $C$ is defined as the ratio $a_n/a_{n+2}$ of two successive expansion coefficients in the limit $n\to\infty$, 
\beq\label{Cb}
C\equiv \lim_{n\to\infty}\0{a_n}{a_{n+2}}\ .
\eeq
This limit is computed as follows. First consider the definition \eq{ak} and notice that the term $P^{-n}$ will strongly suppress $a_n$ in the limit $n\to\infty$ because $P^2$ is strictly positive and diverging for large momenta (see Fig.~1). The sole contribution to the integrand will then come from the minimum of $P^2$ where the integrand is the least suppressed. Second, taking into account that the kernel $K$ is well-behaved, we can conclude that the radius of convergence is given by the 'saddle point approximation' to \eq{Cb}, namely
\beq \label{Cb-max}
C=\min_{y\ge 0}\, P^2(y)\ .
\eeq 
The radius of convergence does not depend on the kernel $K$, but still on the parameters of the coarse-graining. The optimisation criterion for amplitude expansions $C_{\rm opt}=\max_{\rm RS} C$ coincides with the optimisation condition introduced earlier. The finiteness of $C$ is due to the pole structure of threshold functions. This implies that the radius of convergence cannot be increased by changing to another expansion point. The optimal expansion point is $\omega =0$.
 
The optimal coarse-grainings have another interesting interpretation. Consider again the definition \eq{ak} for the expansion coefficients, and let us split them as $a_n=a^<_n+a^>_n$ into its low momentum contributions $a^<_n=\int_0^1 dy K[r]P^{-n}$  and its high momentum contributions $a^>_n=\int_1^\infty dy K[r]P^{-n}$. We assume that the coarse-graining scheme is parametrised by some parameter $b$. Let us define implicitly the parameters $b_n$ from the requirement that the hard modes and soft modes give equal contributions to the coefficients $a_n$, hence $a^>_n=a^<_n$. This is certainly possible for $n$ sufficiently large. The optimal choice for the coarse-graining $b_{\rm opt}$, as defined by the optimisation condition, can be shown to correspond precisely to the limit $b_{\rm opt}=\lim_{n\to\infty} b_n$.
\\[3ex] 
\noindent 
{\bf Field amplitude expansions}\\[-1ex] 

It is worth pointing out a crucial difference between expansions based on the amplitudes $\omega_i$, as considered here, and expansions based directly on {\it field} amplitudes $\phi^a$ or $\rho$ (sometimes also denoted as a local polynomial approximation), which have often been used in the literature \cite{Tetradis:1994ts,MorrisSharp,AbelHiggs3d,Aoki,LPS}. 

We consider the $N$-component scalar field theory with the flow equation \eq{example} as example. Expansions in field amplitudes $\rho$ of the effective potential read $u'=\sum^M_{n=1} \s01{n!}\lambda_n (\rho-\lambda_0)^n$, where $\lambda_0\neq 0$ corresponds to the chosen expansion point.\footnote{For expansions about $\rho=0$ we write $u'=\sum^M_{n=0} \s01{n!}\lambda_n \rho^n$ instead.} The partial differential equation \eq{example} reduces to a set of coupled ordinary differential equations $\beta_n\equiv d\lambda_n/dt$ for the $(M+1)$ couplings $\lambda_n$. These $\beta$-functions depend on the coarse-graining only through the threshold function $\ell^d_1(\omega)$ and derivatives thereof, evaluated at the expansion point $\omega(\lambda_0)$. The relevant amplitudes are $\omega_1=u'$ and $\omega_2=u'+2\rho u''$, and the optimal expansion point is $\omega_1=\omega_2=0$. When expressed in terms of field amplitudes $\rho$, these expansion points are {\it not} equivalent.\footnote{The condition $\omega_1=\omega_2=0$ over-constrains the choice for an expansion point $\lambda_0$ as it implies either $u'=u''=0$, or $u'(0)=0$ and $u''(0)=$ finite.} An expansion in $\rho$ therefore has a smaller radius of convergence than expansions in the amplitudes $\omega_i$. 

Interestingly, there are two exceptions in the present example. The first one is the limit $N\to\infty$, in which case only the amplitude $\omega_1=u'$ remains. The expansion about the local minimum of the scaling potential $\rho_1$ with $\omega_1(\rho_1)=0$ is the optimal choice and equivalent to the expansion in the amplitude $\omega_1$. A second exceptional case concerns $N=1$, where only the amplitude $\omega_2=u'+2\rho u''$ appears, suggesting that an expansion about the inflection point $\rho_2$ with $\omega_2(\rho_2)=0$ has best convergence properties, as long as $\omega$ is Taylor-expandable about $\rho_2$. 
 
It was already argued that expansions about the local potential minimum $\rho_1$ show better convergence than expansions about vanishing field $\lambda_0=0$ \cite{Aoki}. The present discussion clarifies from a more general perspective why an appropriately chosen field expansion point does improve the convergence of the entire series. While the optimal expansion point for amplitude expansions is always $\omega_i=0$, the optimal expansion point for field amplitudes $\rho_{\rm opt}$ depends on the particular theory studied (in the example discussed here on the number of real scalar fields $N$). Clearly, all these expansions, either in $\rho$ or in the amplitudes $\omega_i$, are optimised through an appropriate choice of the coarse-graining, as defined in \eq{opt}.
\\[3ex]
\noindent
{\bf Applications}\\[-1ex]

We now turn our attention to a number of specific coarse-graining functions. Our list of regulators is by no means exhaustive, though it covers the main regulators used and discussed in the literature.  It includes exponential, algebraic and mass-like ones, and the sharp cut-off for comparison (see Tab.~2; the parameter $c$ ensures the consistent normalisation of the scale parameter $k$). They all differ in the way how high momentum modes are cut off, with the parameter $b$ measuring the relative `smoothness' -- the smaller $b$ the `smoother' the coarse-graining. 

\begin{center}
\begin{tabular}{llccccc}
\hline\hline\\[-1ex]
\multicolumn{1}{c}{Regulator} & \multicolumn{2}{c}{Parameter range} &Class& $\quad b_{\rm opt}\quad $&$C_{\rm opt}$ 
\\[.5ex] \hline\\[-1.5ex] 
$r_{\rm exp}\ $ = $\left[\exp cy^b -1\right]^{-1}$
& $ c=\ln 2$
& $1\le b\le \infty$ 
&II${}^*$
&$1.44$
&2
\\[1ex] 
$r_{\rm mexp} $ = $b\left[\exp cy -1\right]^{-1}$
& $ c=\ln (1+b)\ $
& $0\le b\le \infty$ 
&I
&$3.92$
&2
\\[1ex] 
$r_{\rm mod}\, $ = $\left[\exp \s0cb (y+(b-1)y^b)-1\right]^{-1}\ $
&$ c=\ln 2$
&$1\le b\le \infty$
&I
&$1.92$
&2
\\[1ex] 
$r_{\rm mix}\ $ = $\exp \left[-\s0{b}{2a}(y^a-y^{-a})\right]$
&$ a\ge 0$
&$1\le b\le \infty$
&II${}\ $
&2
&2
\\[1ex] 
$r_{\rm step}\ $ = $\s0{2b-2}{b}\,y^{b-2}\left[\exp c\,y^{b-1}-1\right]^{-1}$
&$ c=\ln\0{3b-2}{b}$
&$1\le b\le \infty$
&I
&$4.40$
&$1.65$
\\[1ex] 
$r_{\rm power}$ = $y^{-b}$
& 
&$1\le b\le \infty$
&II${}^*$
&2
&2
\\[1ex] 
$r_{\rm sharp}$ = $\01{\theta(y-1)}-1$
& 
&  
&II${}\ $
& 
& 1
\\[1ex]
\hline\hline
\end{tabular}
\vskip.5cm
\begin{minipage}{.92\hsize}
{\label{Tab2}\small Tab.~2: The definition of various classes of regulator functions and their optimal parameter choices, including the sharp cutoff for comparison. Class II${}^*$ denotes Class II regulators except for $b=1$, where they correspond to Class I.}
\end{minipage} 
\end{center}
${}$\\[-1ex] 

More specifically, the regulators  $r_{\rm exp}, r_{\rm mod}, r_{\rm step}$, $r_{\rm mexp}$ for $b\neq 0$ and $r_{\rm mix}$ for $a\neq 1$ are exponential ones, that is, exponentially suppressed for large momenta. In contrast, $r_{\rm power}$, $r_{\rm mexp}$ for $b=0$, and $r_{\rm mix}$ for $a=0$ are algebraic regulators.  In the classifaction of Tab.~1, the mass-like exponential regulator $r_{\rm mexp}$, the modified exponential regulator $r_{\rm mod}$ and $r_{\rm step}$ are Class I (mass-like) regulators, as are the exponential one $r_{\rm exp}$ and the algebraic one $r_{\rm power}$ for $b=1$. On the other hand, the mixed exponential regulator $r_{\rm mix}$, the sharp cut-off $r_{\rm sharp}$, and $r_{\rm exp}$ and $r_{\rm power}$ for $b>1$, are Class II regulators.

Some cross-dependences amongst these functions are worth noticing. For $b\to\infty$, $r_{\rm exp}, r_{\rm mexp}, r_{\rm mod}$ and $r_{\rm power}$ approach the sharp cut-off limit $r_{\rm sharp}$, while $r_{\rm step}$ approaches $\approx y^{-1}\theta(1-y)$, which corresponds to a mass term which is cut-off for momenta above $k$, $R_{{\rm step},k}\approx k^2 \theta(k^2-q^2)$. The two-parameter family of mixed exponential regulators $r_{\rm mix}$, which obey $r_{\rm mix}(1/y)=1/r_{\rm mix}(y)$, contain the algebraic regulator $r_{\rm power}$ as the limiting case $a\to 0$. $r_{\rm exp}$ and $r_{\rm mod}$ for $b=1$ coincide with $r_{\rm step}$ for $b=2$. Also, $r_{\rm mexp}$ for $b=0$, and $r_{\rm step}$ for $b=1$ correspond to $r_{\rm power}$ for $b=1$. 
 
We now distinguish the different coarse-grainings by their associated radii of convergence. In Fig.~2, we have displayed $C$, as defined in \eq{Cb}, for different coarse-graining schemes and as functions of the coarse-graining parameter $b$. In the sharp cutoff limit $b\to\infty$ the value $C_{\rm sharp}=1$ is attained. For the classes of regulators considered here, we notice that the maximal radius is reached for specific smooth coarse-grainings. The radii vary between $1\le C\le 2$ (though it is possible to find regulators such that $C$ is arbitrarily small). Optimal parameters are found for both Class I and Class II regulators (see Fig.~3). The most efficient way for determining optimal parameters for all Class Ib and Class II regulators is the following. We have established that $P^2(y)$ takes its maximum at $y=1$ and $r=1$ for an optimal coarse-graining. The vanishing of $dP^2/dy=1+r+yr'$ at the extremum implies $r'_{\rm opt}(1)=-2$. This allows to fix the optimal coarse-graining parameters.
\\

\begin{figure}[t]
\begin{center}
\unitlength0.001\hsize
\begin{picture}(1000,500)
\put(250,180){
\begin{tabular}{ll}
$r_{\rm power}$  &$ {}^{\put(0,0){\line(80,0){70}}}${}\\[-.5ex]
$r_{\rm exp}$ &$ {}^{\multiput(0,0)(20,0){4}{\line(10,0){10}}} $\\[-.5ex]
$r_{\rm mod}$ &$ {}^{\multiput(0,0)(20,0){3}{\put(0,0){\line(10,0){10}}\put(14,0){\line(2,0){2}}}\put(60,0){\line(10,0){10}}}${}
\end{tabular}}
\put(750,200){
\begin{tabular}{ll}
$r_{\rm mix}$&\\
\footnotesize $a=0$  &$ {}^{\put(0,0){\line(80,0){70}}}${}\\[-.5ex]
\footnotesize $a=\s012$&$ {}^{\multiput(0,0)(10,0){8}{\line(5,0){5}}}${}\\[-.5ex]
\footnotesize $a=1$    &$ {}^{\multiput(0,0)(20,0){3}{\put(0,0){\line(10,0){10}}\put(14,0)
               {\line(2,0){2}}}\put(60,0){\line(10,0){10}}}${}\\[-.5ex]
\footnotesize $a=2$    &$ {}^{\multiput(0,0)(20,0){4}{\line(10,0){10}}} $
\end{tabular}}
\put(650,300){\framebox{\large $C=\min_{y\ge 0} P^2(y)$}}
\put(260,70){\Large $b$}
\put(730,70){\Large $b$}
\hskip.05\hsize
\psfig{file=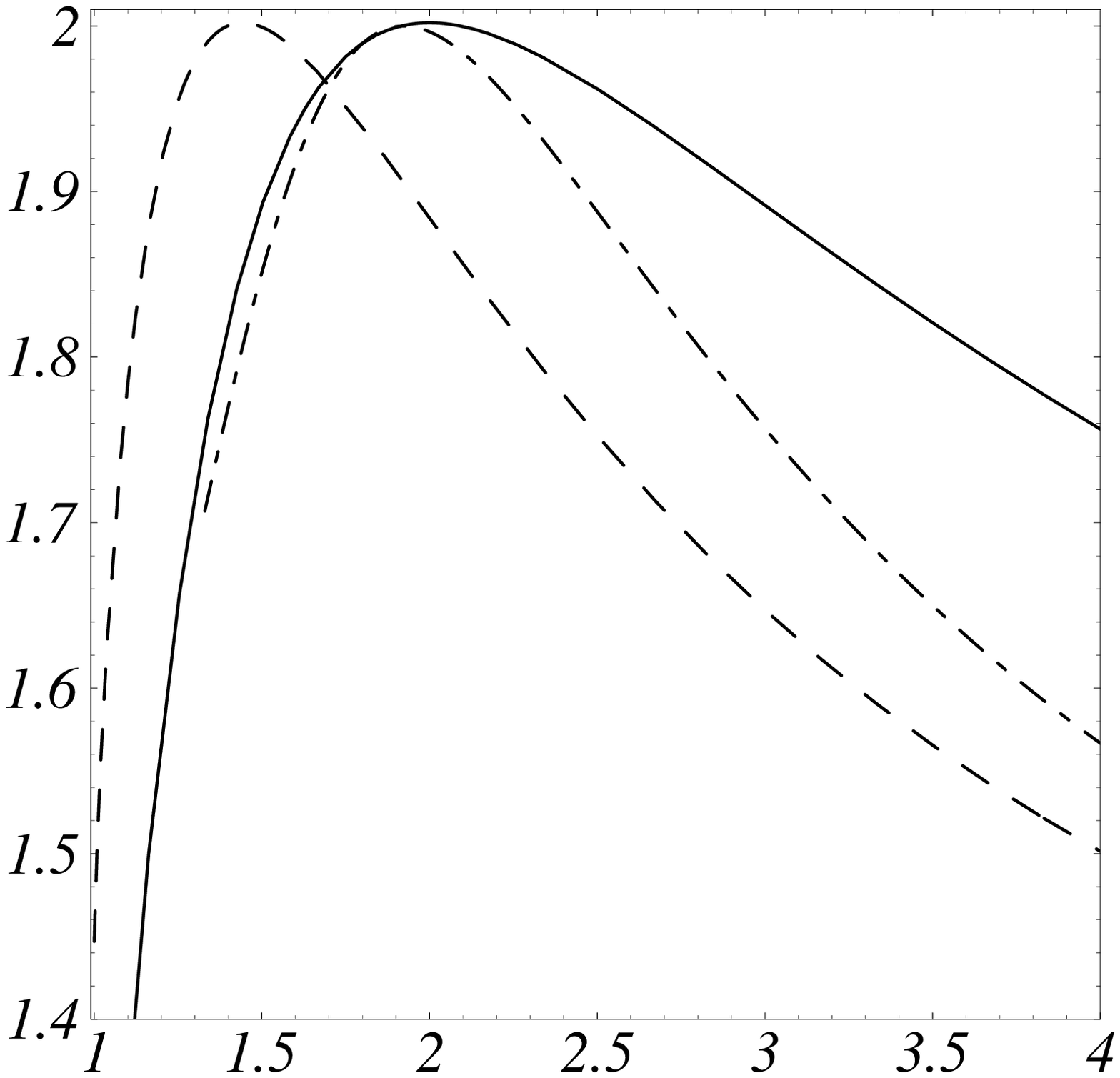,width=.45\hsize}
\hskip-.01\hsize
\psfig{file=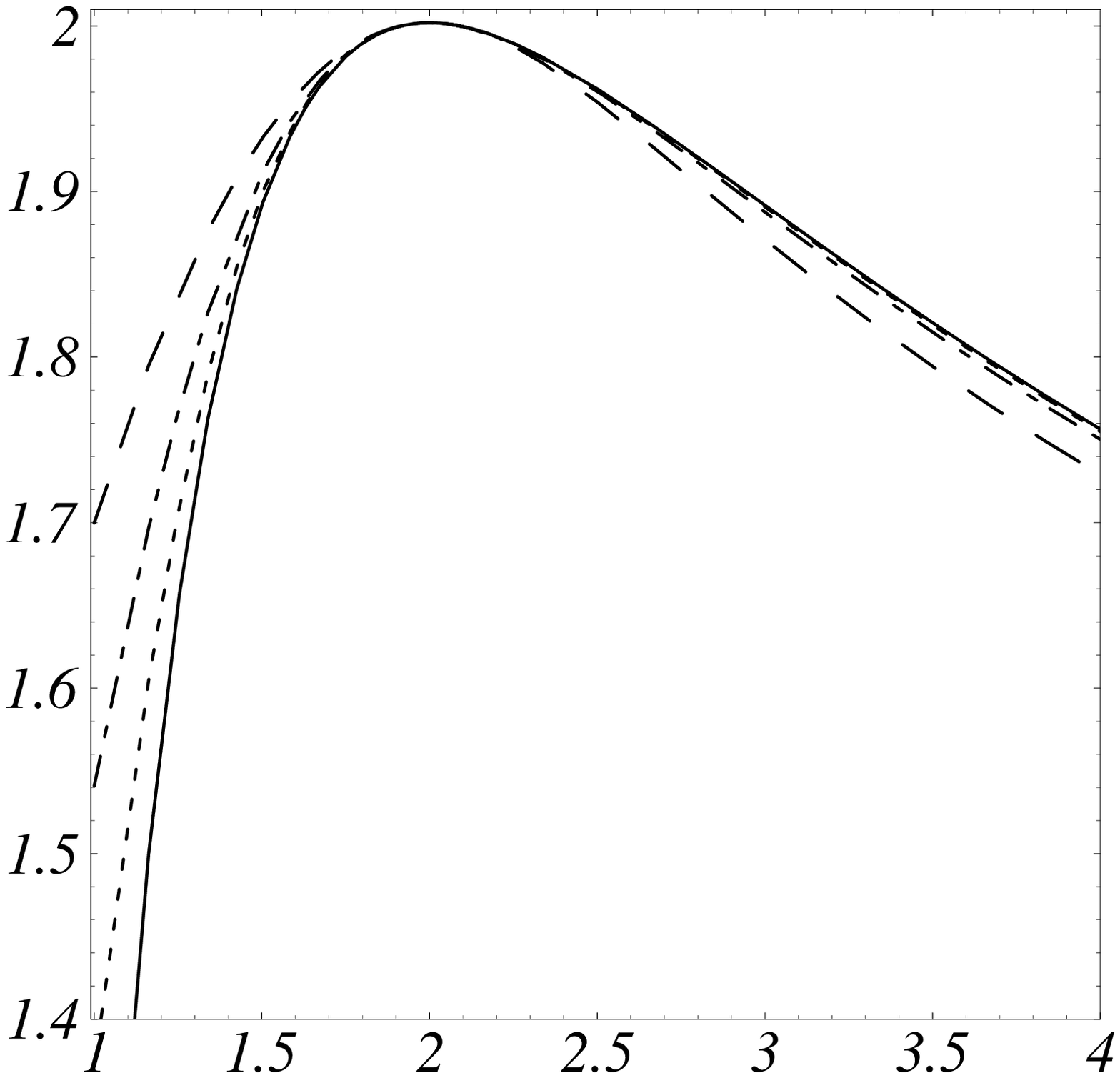,width=.45\hsize}
\end{picture}
\vskip-.5cm \begin{minipage}{.92\hsize}
{\small Fig.~2: The dimensionless radii of convergence for amplitude expansions ({\it a.k.a.}~location of the pole for threshold functions) as a function of the coarse-graining parameter for the regulators $r_{\rm power}$, $r_{\rm exp}$ and $r_{\rm mod}$ (left panel) and $r_{\rm mix}$ with $a=0,\s012,1$ and $2$ (right panel).}
\end{minipage} 
\end{center}
\end{figure}

For $r_{\rm exp}$ one finds $b_{\rm exp}=1/\ln 2$, for $r_{\rm power}$ the radius of convergence reads $C=b(1-b)^{1/b-1}$ which attains its maximum at $b_{\rm power}=2$. The optimal parameter for $r_{\rm mix}$ is $b_{\rm mix}=2$ for all $a$. The value for $b_{\rm mexp}$ is obtained numerically by solving $2b=(b+1)\ln(b+1)$, and the one for $b_{\rm mod}$ reads $b_{\rm mod}=[1+\0{1}{\ln 2}] [\012+\sqrt{\014-1/(1+\0{1}{\ln 2})^2}]$ (see Tab.~2). The regulator $r_{\rm step}$ does not attain the maximal radius $C=2$. The reason behind it is that $P^2_{\rm step}(y)$ has two local minima for $b \approx b_{\rm opt}$, one at $y\approx 1$, and another one at $y=0$.  The minima are degenerate for $b=b_{\rm opt}$, and their competition only allows for a sub-optimal value $C_{\rm step}=1.65$.

The fact that $C_{\rm sharp}=1$ is half as big as the possible optimal value explains why amplitude expansions of sharp cut-off flows have a notoriously bad convergence behaviour \cite{MorrisSharp}. This property is not a particularity of the sharp cut-off; the smooth cut-offs $r_{\rm power}$, $r_{\rm exp}$ and $r_{\rm mexp}$ (all for $b=1$) also have a radius of convergence $C=1$. A significant improvement is achieved only for the optimal coarse-grainings, which all belong to smooth cut-offs.

\begin{figure}[t]
\begin{center}
\unitlength0.001\hsize
\begin{picture}(500,500)
\put(180,350){\framebox{\large $P_{\rm opt}^2(q^2/k^2)$}}
\put(200,-20){\large ${q^2}/{k^2}$}
\psfig{file=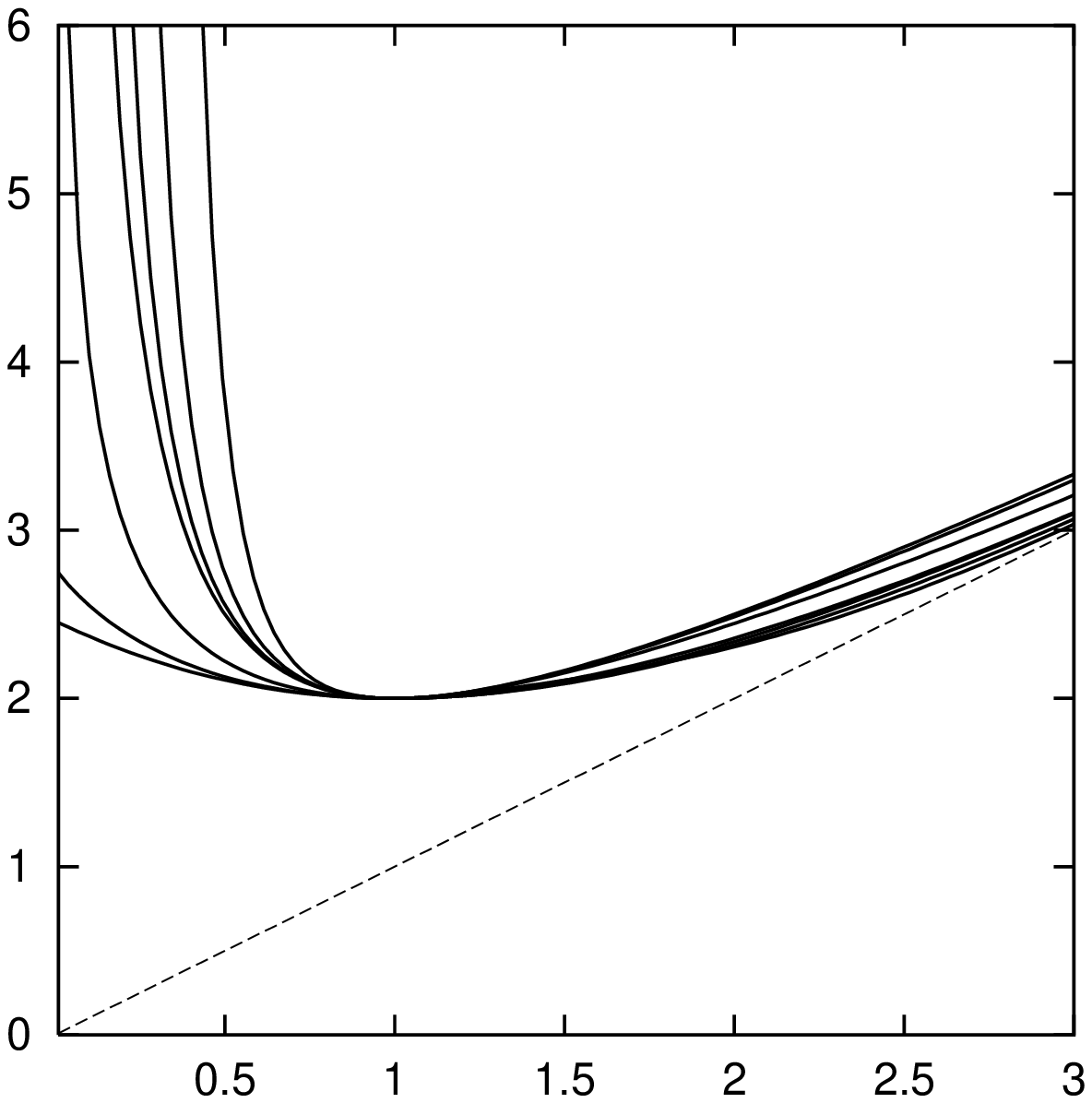,width=.45\hsize}
\end{picture}
\vskip.8cm
\begin{minipage}{.92\hsize}
{\small Fig.~3: The optimised effective inverse propagators for different regulators (thick full lines). From the left-lowest curve clockwise: $r_{\rm mexp}$, $r_{\rm mod}$, $r_{\rm exp}$, $r_{\rm power}$, $r_{\rm mix}(a=\s012)$, $r_{\rm mix}(a=1)$ and $r_{\rm mix}(a=2)$. The dashed line corresponds to $r=0$.}
\end{minipage} 
\end{center}
\end{figure}

The present discussion can be extended in several directions. Fermionic fields can be included using a regulator term like  $\sim \int d^dq\bar \psi(-q)\,R_{{\rm F},k}(q)\,\psi(q)$, with $R^2_{{\rm F},k}(q)=q^2\, r^2_{\rm F}(q^2/k^2)$. Effectively, it appears in threshold functions in the combination $P^2_{\rm F}=y[1+r_{\rm F}(y)]^2$, which is the fermionic analogue of \eq{P2}. Hence, the optimisation condition for fermionic degrees of freedom is the same as \eq{opt}, replacing $P^2$ by $P^2_{\rm F}$, yielding $C_{\rm F,opt}=4$. For gauge fields, the flow equation is amended by a (modified) Ward or a (modified) BRST Identity, which ensures that physical Green functions obey the usual Ward or BRST Identities \cite{gaugefields}. The criterion \eq{P2} is compatible with such an additional constraint. 
 Also, wave function renormalisations can be taken into account in the usual manner, setting $R_k=Z_k q^2 r(q^2/k^2)$. This modification will not change the optimisation as discussed in the present note, and allows for a systematic inclusion of higher derivative operators. The same holds for field theories at finite temperature within the imaginary or the real-time formalism \cite{Litim98}.
 
In conclusion, we have presented a simple and generic optimisation criterion for exact renormalisation group equations. 
It is interesting to mention its close link to a minimisation of the spurious scheme dependence of flow equations, a more detailed discussion of which shall be presented elsewhere \cite{elsewhere}.\\

The author thanks Ulrich Ellwanger, Filipe Freire, Jan M.~Pawlowski, Janos Polonyi and Christof Wetterich for useful discussions.


\def\PRA#1#2#3#4#5{ #1,\,\,{\it }\,Phys.\,Rev.\,{\bf A#3}\,(19#4)\,#5}
\def\PRB#1#2#3#4#5{#1,{\it }\,Phys.\,Rev.\,{\bf B#3}\,(19#4)\,#5}
\def\PRL#1#2#3#4#5{#1,\,\,{\it }\,Phys.\,Rev.\,Lett.\,{\bf #3} (19#4) #5}
\def\PRC#1#2#3#4#5{#1,\,\,{\it }\,Phys.\,Rev.\,{\bf C#3}\,(19#4)\,#5}
\def\PRD#1#2#3#4#5{#1,{\it }\,Phys.\,Rev.\,{\bf D#3}\,(19#4)\,#5}
\def\PRE#1#2#3#4#5{#1,\,\,{\it }\,Phys.\,Rev.\,{\bf E\,#3}\,(19#4)\,#5}
\def\PRep#1#2#3#4#5{#1,{\it }\,Phys.\,Rep.\,{\bf  #3}\,(19#4)\,#5}
\def\NPB#1#2#3#4#5{#1,{\it }\,Nucl.\,Phys.\,{\bf B#3}\,(19#4)\,#5}
\def\PLB#1#2#3#4#5{#1,{\it }\,Phys.\,Lett.\,{\bf B#3}\,(19#4)\,#5}
\def\JPA#1#2#3#4#5{#1,\,\,{\it }\,J.\,Phys.\,{\bf A#3}\,(19#4)\,#5}
\def\JPB#1#2#3#4#5{#1,\,\,{\it }\,J.\,Phys.\,{\bf B#3}\,(19#4)\,#5}
\def\JPC#1#2#3#4#5{#1,\,\,{\it }\,J.\,Phys.\,{\bf C#3}\,(19#4)\,#5}
\def\ZPC#1#2#3#4#5{#1,{\it }\,Z.\,Phys.\,{\bf C#3}\,(19#4)\,#5}
\def\MPLA#1#2#3#4#5{#1,{\it }\,Mod.\,Phys.\,Lett.\,{\bf A#3}\,(19#4)\,#5}
\def\and#1#2#3{{\bf #1}\,(19#2)\,#3}

\end{document}